\documentclass[aps,prl,amsmath,amssymb,amsfonts,floatfix,nofootinbib,superscriptaddress,longbibliography,12pt,reprint,noeprint]{revtex4-2}
\usepackage[utf8]{inputenc}
\usepackage[T1]{fontenc}
\usepackage{amsmath}
\usepackage{amssymb}
\usepackage{physics}
\usepackage{mathtools}
\usepackage{array}
\usepackage{graphicx}
\usepackage[dvipsnames]{xcolor}
    \definecolor{darkgreen}{rgb}{0,0.5,0}
    \definecolor{darkblue}{rgb}{0,0,0.6}
    \definecolor{purple}{rgb}{0.4,.2,0.7}
	\definecolor{linkblue}{rgb}{0.0,0.0,0.6}
	\definecolor{citepurple}{rgb}{0.35,0.2,0.55}
\usepackage{tikz}
\usepackage{relsize}
\usepackage{CJKutf8}
\usepackage[hyperfootnotes = false, 
	colorlinks = true,
	linkcolor = darkblue, 
	citecolor = purple, 
	urlcolor = linkblue]{hyperref}

\newcommand{\ii}{\mathrm{i}}
\newcommand{\ee}{e}
\newcommand{\sgn}{\operatorname{sgn}}

\makeatletter
\newcommand{\startendmatter}{%
  \close@column@grid
  \clearpage
  \twocolumngrid
}
\makeatother

\begin{document}

\title{Tuning quantum magic of pure quantum chaotic states with a gravity dual}
\author{Antonio M. Garc\'ia-Garc\'ia}
\email{amgg@sjtu.edu.cn}
\affiliation{Shanghai Center for Complex Physics, School of Physics and Astronomy,
	Shanghai Jiao Tong University, Shanghai 200240, China
}
\author{Xianlong Liu ({\begin{CJK}{UTF8}{gbsn}刘显龙\end{CJK}})}
\email{xianlong\_liu@sjtu.edu.cn}
\affiliation{Shanghai Center for Complex Physics, School of Physics and Astronomy,
    Shanghai Jiao Tong University, Shanghai 200240, China
}

\author{Jie-ping Zheng ({\begin{CJK*}{UTF8}{gbsn}郑杰平\end{CJK*}})}
\email{jpzheng@sjtu.edu.cn}
\affiliation{Shanghai Center for Complex Physics, School of Physics and Astronomy, 
    Shanghai Jiao Tong University, Shanghai 200240, China
}

\begin{abstract}
Quantum magic is a fundamental resource that quantifies to what extent quantum states can be efficiently simulated on a classical computer. We study it for states constructed from the Sachdev-Ye-Kitaev (SYK) Hamiltonian with $N$ Majoranas by the fermionic anti-flatness (FAF). We show analytically that, in the large $N$ limit, the quantum magic of pure Kourkoulou-Maldacena (KM) states, dual to a quantum black hole with an end-of-world particle behind the horizon, is linear in $N$ with a slope, depending on the black hole temperature, that can be tuned between zero and $1/2$. By contrast, the FAF of Gaussian states evolved in real time with the SYK Hamitonian approaches $\approx N/2$ exponentially at a rate given by a multiple of the leading Ruelle-Pollicot resonance. Subleading corrections in $N$ for SYK energy eigenstates, computed numerically for $N \leq 54$ by combining Krylov subspace with GPU acceleration techniques, decay exponentially with $N$, but power-law if the SYK couplings are sparsified, and are order of magnitude larger for states close to the ground state, a region with an established gravity analogue. Our results offer new insights about the relation between quantum information, quantum chaos and low-dimension quantum gravity.

\end{abstract}

\maketitle

Quantum magic \cite{kitaev2005,veitch2012,veitch2014,chitambar2019} quantifies the departure of a quantum state from classes of states that can be efficiently simulated on a classical computer, such as Gaussian states. In the context of quantum information, quantum magic measures the non-stabilizerness of the state, which can be understood as its deviation from a Gaussian state.
Therefore, the state magic content reflects its quantum  complexity, an aspect of its {\it quantumness}, that complements others such as entanglement that probes quantum non-locality. 

Observables introduced to characterize magic includes Wigner negativity \cite{veitch2012,pashayan2015}, mana \cite{veitch2014,zhang2024}, thauma \cite{wang2020}, and more recently the stabilizer R\'enyi entropy (SRE) \cite{leone2022}. The latter is monotonic in magic \cite{leone2024} and typically easier to compute than the other observables, which has made possible its study in a variety of contexts: quantum many-body systems \cite{liu2022}, conformal field theory \cite{hoshino2026}, traversable wormholes \cite{pengfei2026} and other gravity settings \cite{cao2025}, and quantum circuits \cite{turkeshi2025,bejan2025,sierant2026a}. Quantum magic has already been measured experimentally \cite{niroula2024} and can also be estimated efficiently in quantum computers~\cite{haug2023}. 
 
Despite these recent advances ~\cite{ding2025,sierant2026}, the evaluation of magic is still challenging in a many-body context. The recent introduction of fermionic anti-flatness~(FAF) \cite{sierant2026b,robin2025,falcao2026,tirrito2024} has further simplified its calculation. for $N$ Majorana fermions $\chi_i$, $i=1,\dots,N$, obeying $\{\chi_i,\chi_j\}=\delta_{ij}$, the FAF of a state $\ket{\Psi}$ is defined by \cite{sierant2026b}
\begin{equation} \label{eq:FAF_def}
	\mathcal{F}_{k}(\ket{\Psi}) = \frac{N}{2} - \tr[(M^{\rm{T}} M)^k] ,
\end{equation}
where $M_{mn} \equiv - \ii \bra{\Psi}[\chi_m, \chi_n]\ket{\Psi}$. It captures the sharp distinction between Gaussian states, where it vanishes, and sufficiently complex states, such as Haar-random states, where it is possible to show analytically~\cite{sierant2026b} that it becomes extensive $\mathcal{F}\sim N/2$ for $N \to \infty$  with $N/2$ the logarithm of the  Hilbert space dimension. 

In this paper, we study the FAF of states generated from the Sachdev-Ye-Kitaev (SYK) model \cite{kitaev2015,french1970,bohigas1971,sachdev1993,benet2001,garcia2016,maldacena2016,garcia2022d} that describes $N$ Majorana fermions with random all-to-all interactions. The combination of analytical tractability \cite{maldacena2016,kitaev2015}, quantum chaotic dynamics \cite{kitaev2015,garcia2016}, and the existence of a gravity dual \cite{maldacena2016b} makes it an ideal laboratory to characterize quantum magic in strongly interacting quantum many-body systems with intriguing implications for low dimensional quantum gravity. Quantum magic, through the calculation of the SRE, has already been studied numerically in the SYK model~\cite{bera2025,jasser2025,russomanno2025}. For instance in Ref.~\cite{bera2025}, it was shown numerically that chaotic and not chaotic SYK's models share the same scaling with system size but the chaotic one has a larger magic. A two-site SYK model, dual to a traversable wormhole \cite{maldacena2018}, has been shown \cite{pengfei2026,bettaque2026} to undergo a transition in magic. The quantum magic of a thermo-field double state has been related \cite{pengfei2026b} to the spectral form factor, which provides an interesting relation between quantum magic and quantum chaos.

Here, we employ the FAF to characterize quantum magic in several classes of states associated with the SYK Hamiltonian. We study in particular the FAF of Kourkoulou-Maldacena (KM) pure states~\cite{kour2017,pengfei2020}, a one parameter family of states resulting from the Euclidean evolution of Gaussian states with the SYK Hamiltonian, and SYK energy eigenstates. We also investigate the real-time evolution of FAF for Gaussian initial states. We start with the definition of FAF and then introduce the states considered below. 

\emph{The model and definition of KM pure states.--} We consider the quartic ($q = 4$) SYK Hamiltonian for $N$ Majorana fermions $\chi_i$, 
\begin{equation} \label{eq:sykq4}
	H_{\rm SYK} = \sum_{i<j<k<l} J_{ijkl}\,\chi_i\chi_j\chi_k\chi_l	
\end{equation}
with $\overline{J_{ijkl}}=0$ and $\overline{J_{ijkl}^2}=6J^2/N^3$. The large $q$ case is discussed in the End Matter.
The SYK model is strongly interacting, quantum chaotic at all time scales~\cite{kitaev2015,garcia2016,garcia2026a}, analytically tractable in different ranges of parameters~\cite{kitaev2015,maldacena2016} and has a gravity dual~\cite{kitaev2015,maldacena2016b}. The pure KM states \cite{kour2017}, which constitute one of our main interests, are constructed from the SYK model Eq.~\eqref{eq:sykq4} as follows. We first introduce the mutually commuting operators
$	\hat S_k=-2\ii\,\chi_{2k-1}\chi_{2k}$ with $k=1,\ldots,N/2$,
whose simultaneous eigenstates $\ket{B_{\vec s}}$ satisfy
\begin{equation}
	\hat S_k\ket{B_{\vec s}}=s_k\ket{B_{\vec s}},\qquad s_k=\pm1.
\end{equation}
These states are Gaussian product states in the paired Majorana basis. Their Euclidean evolution under the SYK Hamiltonian defines the normalized pure KM states
\begin{equation}
	\ket{B_{\vec s}(\beta)}=\frac{e^{-\beta H_{\rm SYK}/2}\ket{B_{\vec s}}}
	{\sqrt{\bra{B_{\vec s}}e^{-\beta H_{\rm SYK}}\ket{B_{\vec s}}}},
\end{equation}
which interpolate continuously between the simple Gaussian states $\ket{B_{\vec s}}$ at $\beta=0$ and strongly correlated low-energy states as $\beta$ increases. As we shall see, this construction allows us to tune the amount of fermionic magic with an effective temperature parameter. The original motivation for introducing KM states is their gravitational interpretation. In the large $N$ limit, they are dual to near AdS$_2$ quantum black holes with an end-of-the-world particle behind the horizon \cite{kour2017}, which can be made visible to a boundary observer by perturbing the SYK model with operators $\hat S_k$, using an $s_k$ configuration equal to that used to construct the KM state. Our results therefore have a direct relevance in this context.

\emph{Analytical FAF for KM states and time evolution.--}
We consider the $k = 1$ FAF $\mathcal{F} \equiv \mathcal{F}_{k=1}$, with $\mathcal{F}_{k}$ defined in Eq.~\eqref{eq:FAF_def}, for the pure states $\ket{B_{\vec{s}}(\beta)}$. In large $N$ limit,  the theory exhibits an emergent $\mathrm{O}(N)$ symmetry, which contains the Flip Group \cite{kour2017} as a discrete subgroup, whose generators flip individual components of $\vec{s}$. This selection rule implies that the only non-vanishing matrix elements entering the FAF are $\bra{B_{\vec{s}}(\beta)} \chi_{2i-1} \chi_{2i} \ket{B_{\vec{s}}(\beta)}$, $i=1,2,\ldots,N/2$, referred to as ``off-diagonal'' correlators~\cite{kour2017}. As a consequence, the FAF reduces to
\begin{equation}\label{eq:fafbana}
	\mathcal{F}(\beta) \equiv \mathcal{F}(\ket{B_{\vec{s}}(\beta)}) =  N \Biggl[
	\frac{1}{2} - 8 G_{\rm{E}}\biggl(\frac{\beta}{2}\biggr)^4
	\Biggr] ,
\end{equation}
where $G_{\rm{E}}(\beta/2)$ is the value at $\tau = \beta / 2$ of the Euclidean Green's function  $G_{\rm{E}}(\tau) \equiv \sum_{i=1}^{N} \langle \operatorname{T} \chi_i(\tau) \chi_i(0) \rangle_{\beta} / N$ given by the solution of the large $N$ Schwinger-Dyson (SD) equations~\cite{maldacena2016}. In the large $N$ limit, this relation is independent of $\vec{s}$ and therefore holds for all such pure KM states. More details of the derivation can be found in the End Matter. The real time evolution of FAF follows by analytic continuation
\begin{equation}\label{eq:faftana}
	\mathcal{F}(t; \beta) = N  \biggl[
	\frac{1}{2} - 8 G^{\rm{W}}(t)^4
	\biggr] ,
\end{equation}
where $G^{\rm{W}}(t) \!\! \equiv \!\! ( \! N \! Z_{\beta} \! )^{-1}\!\sum_{i=1}^{N} \!\! \operatorname{tr}[\ee^{-\beta H \!/ 2} \chi_i(t) \ee^{-\beta H \!/ 2} \chi_i(0))]$ denotes the regularized Wightman correlation function, or equivalently, $G^{\rm{W}}(t) \equiv G_{\rm{E}}(\beta / 2 + \ii t)$~\cite{maldacena2016}. The $1/N$-correction might be computed from the time-ordered connected four-point function of the SYK model~\cite{maldacena2016}.
\begin{figure}[!tb]
  \centering
  \begin{tikzpicture}
    \node[inner sep=0pt, outer sep=0pt] (panelA) {%
      \includegraphics[width=\columnwidth]{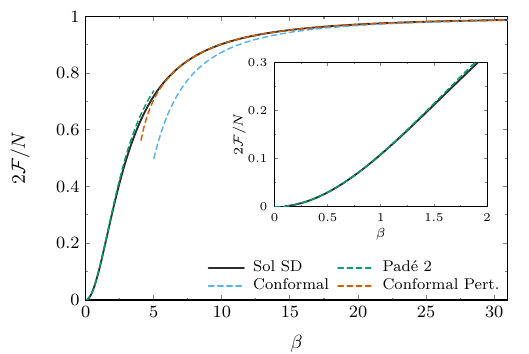}%
    };
    \node[
      anchor=north west,
      inner sep=0pt,
      outer sep=0pt,
      xshift=1pt,
      yshift=-1pt,
      font=\fontsize{9}{10.5}\selectfont\bfseries
    ] at (panelA.north west) {(a)};
  \end{tikzpicture}
  \vspace{-1.5mm}
  \begin{tikzpicture}
    \node[inner sep=0pt, outer sep=0pt] (panelB) {%
      \includegraphics[width=\columnwidth]{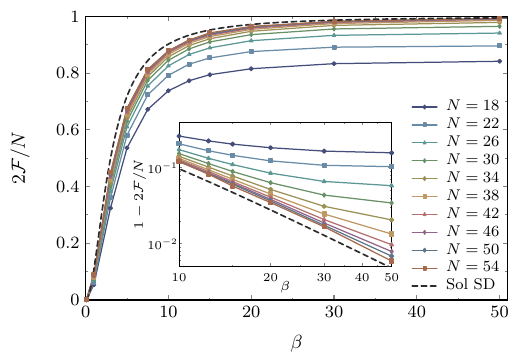}%
    };
    \node[
      anchor=north west,
      inner sep=0pt,
      outer sep=0pt,
      xshift=1pt,
      yshift=-1pt,
      font=\fontsize{9}{10.5}\selectfont\bfseries
    ] at (panelB.north west) {(b)};
  \end{tikzpicture}
  \vspace{-1mm}
  \caption{\textbf{(a)} Analytical $2\mathcal{F} / N$ versus $\beta$ for the pure KM states and $J = 1$. \textbf{(b)} Numerical FAF for the pure KM states as a function of $\beta$ for different $N$'s. The dashed line is the analytical large $N$ result Eq.~(\ref{eq:fafbana}). The agreement between numerical and analytical results is excellent. The inset log-log plot shows that the agreement extends to the large $\beta$ region as well.}
\label{fig:FAF_vs_beta}
\end{figure}

\noindent
\textit{Explicit $\beta$ dependence of KM states.--}
We first consider the FAF of the state $\ket{B_{\vec{s}}(\beta)}$.
In the low temperature limit $\beta J \gg 1$, the Euclidean Green's function admits a systematic perturbative expansion \cite{cruz2023} in $1/\beta J$ around the  conformal limit \cite{maldacena2016} leading to
\begin{equation}
	\mathcal{F}(\beta J) \! = \! N  \Biggl[\!
	\frac{1}{2} 
	- \frac{2\pi}{(\beta J)^2}
	+ \frac{13.311}{(\beta J)^3}
	+ \frac{9.291}{(\beta J)^4}
	+ \frac{38.909}{(\beta J)^{4.77}}
	+ \dots \!
	\Biggr].
\end{equation}
In the $\beta J \ll 1$ limit, we perform a similar expansion,
%\begin{equation}
$	G(\tau) = \frac{1}{2} \sgn(\tau) \sum_{n=0}^{\infty} c_{2n} E_{2n}(|\tau|/\beta)(\beta J)^{2n} $,
%\end{equation}
where $c_{2n}$ are the infinite temperature expansion coefficients \cite{dodelson2025}. The Euler polynomials $E_{n}(x)$ ensure the the KMS condition $G(\tau +\beta) = - G(\tau)$ for any finite $\beta$. Due to the factorial growth, the high-temperature expansion of FAF is asymptotic with vanishing radius of convergence. Nevertheless, the expansion coefficients can be employed to construct the Pad\'{e} approximant \cite{maldacena2016} of $\mathcal{F}((\beta J)^2)$ about $\beta J = 0$. This promotes the truncated series expansion to a rational function and thus provides the pole structure of FAF in the complex $\beta$-plane. The order $[2/2]$ Pad\'{e} approximant gives 
\begin{equation}
	\frac{1}{N}\mathcal{F}_{[2/2]}((\beta J)^2) =
%	\!=\!
	\frac{
		\frac{1}{16} (\beta J)^2 
		+ \frac{1648505}{49222656} (\beta J)^{4} 
	}{ 
		1 
		+ \frac{528293}{769104} (\beta J)^2 
		+ \frac{101178289}{1476679680} (\beta J)^{4} 
	}.
\end{equation}
The higher order approximant does not give better agreement for the high temperature $\beta J \ll 1$ result. In Fig.~\ref{fig:FAF_vs_beta}~(a) we present the results of $2\mathcal{F}(\beta) / N$ based on the solutions of the SYK Schwinger-Dyson equation and the conformal results above. In Fig.~\ref{fig:FAF_vs_beta}~(b), we compare these analytical results with numerical finite $N$ calculations by using exact diagonalization for system sizes $N \leq 26$ and Krylov-subspace/Lanczos techniques together with GPU acceleration for the larger sizes, which enables us to reach up to $N=54$ Majorana fermions.

\begin{figure}[!tb]
\centering
\includegraphics[width=\columnwidth]{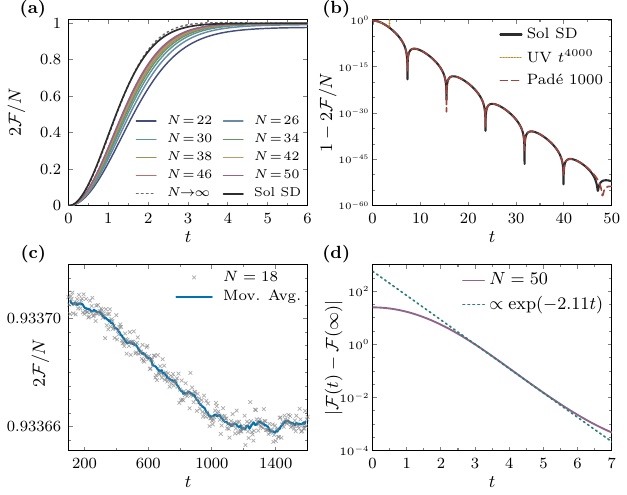}
\caption{
	\textbf{(a)} Time evolution of FAF at finite $N$ with initial state~$\ket{B_{\vec{s}}}$. The black solid curve stands for Eq.~(\ref{eq:faftana}) with~$G^{\rm{W}}(t)$ given by the solution of the SD equations, which agrees well with numerical results ($N \to \infty$) after a linear in $N$ extrapolation. The time scale to reach saturation is largely $N$ independent.
	\textbf{(b)} Analytical time evolution of the normalized FAF with initial state~$\ket{B_{\vec{s}}}$.
	\textbf{(c)} Time evolution of the FAF for $N=18$ around the Heisenberg time, i.e., the inverse of the mean level spacing. The observed ramp-plateau pattern is consistent with Eq.~(\ref{eq:faftana}) and Ref. \cite{garcia2026a,cotler2017}. The number of disorder realizations is $10^7$. 
	\textbf{(d)} The approach of the FAF to its saturation value for $N=50$. The decay rate is close to the analytical prediction \cite{garcia2024,kulkarni2022}.
}
\label{fig:FAF_time_evolution}
\end{figure}

\noindent
\textit{FAF time evolution.--}
We now turn to the explicit calculation of the time dependence of the FAF resulting from an initial Gaussian state $\ket{B_{\vec{s}}}$ that is evolved in real time with the SYK Hamiltonian, in the large $N$ limit,  leading to Eq.~(\ref{eq:faftana}) with $\beta = 0$.  
%For the FAF time evolution Eq with an initial state $\ket{B_{\vec{s}}(\beta)}$, we need the Wightman function. In the conformal limit, we have \cite{maldacena2016},
%\begin{equation}
%$	G^{\rm{W}}_{c}(t) = b \left[ \frac{\pi}{\beta \cosh \frac{\pi t}{\beta} }  \right]^{2 \Delta}.$
%\end{equation}
This corresponds to the infinite temperature limit of $G^{\rm W}$. A short time (UV) expansion \cite{dodelson2025} using the SD equations yields
\begin{equation}
\begin{aligned}
	\frac{1}{N} \mathcal{F}(t) =\;&
	\frac{(J t)^2}{4} 
	-\frac{13 (J t)^4}{192}
	+\frac{307 (J t)^6}{23040}
	-\frac{5549 (J t)^8}{2580480} \\
	& +\frac{141067 (J t)^{10}}{464486400}
	- \frac{957373 (J t)^{12}}{24524881920}
	+ O\bigl((Jt)^{14}\bigr) .
\end{aligned}
\end{equation}
In contrast to the temperature dependence, this expansion has a finite convergence radius approximately equal to $3.56$, rendering it well suited to Pad\'{e} construction. As demonstrated in Fig.~\ref{fig:FAF_time_evolution}~(b), the resulting high order rational approximation extends this short-time expansion beyond its convergence disk, and reproduces the poles responsible for the late-time behavior, exhibiting remarkable agreement with the solution of the SD equations. In Fig.~\ref{fig:FAF_time_evolution} (a), we present the time evolution of the pure states $\ket{B_{\vec{s}}}$ based on the solutions of the SYK${}_4$ SD equation. 

\begin{figure}
	\begin{center}
		\includegraphics[width=\columnwidth]{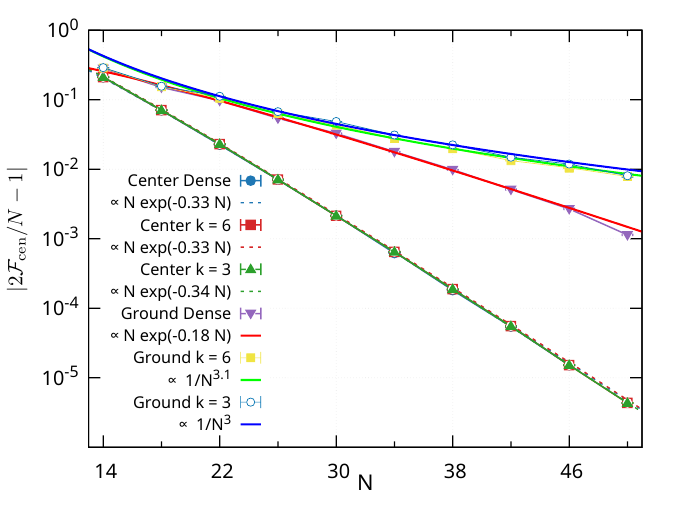}
		\caption{Finite $N$ FAF for the energy eigenstates of the dense and sparse SYK model. Correction to the leading FAF $= N/2$ near the ground state (third excited state) and around the center of the spectrum. The subleading in $N$ FAF is much larger for eigenstates close to the ground state and decays exponentially except for eigenstates of the sparse SYK close to the ground state for which is power-law $\approx 1/N^3$.}
		\label{fig:FAF_groundandcenter}
	\end{center}
\end{figure}

\emph{Numerical FAF beyond the large \texorpdfstring{$N$}{N} limit.--} 
We now turn to the numerical evaluation of the FAF using exact diagonalization for $N \leq 26$ and Krylov subspace technique together with GPU acceleration for $ 26 < N \leq 54$, for confirming the previous analytical large $N$ results, investigating subleading in $N$ corrections and computing FAF for SYK energy eigenstates. In Fig.~\ref{fig:FAF_groundandcenter}, we depict the finite-$N$ correction to the leading $\mathcal{F}\sim N/2$ scaling (not shown) for eigenstates of the SYK model Hamiltonian in the region around the ground state (second excited state from the ground), and also for eigenstates around the center of the spectrum. The subleading correction in $N$ is dramatically larger for states close to the ground state. Indeed, around the center (high energy), the correction is very well described by an exponential decay $\sim N e^{-bN}$  with $b \sim 0.3$, which is qualitatively similar to that of Haar's states \cite{sierant2026a}. This subleading exponential scaling also applies to variants of the SYK model resulting from the sparsification of the SYK random couplings \cite{garcia2021} $J_{ijkl}  \to x_{ijkl}J_{ijkl}$, where $x_{ijkl} = 0$ with probability $1-p$ and $x_{ijkl} = 1$ with probability $p$ provided that $k =\frac{p}{N}\binom{N}{4}\sim O(1)$. This result for the FAF is consistent with that of Ref.~\cite{garcia2021} which showed that the quantum chaotic dynamics is also not substantially altered by sparsing. However, for states close to the ground, although the decay is also exponential in the dense SYK, the exponent is different leading to much larger deviations. Moreover, in this region, if couplings are sparsified, the correction is power-law instead of exponential.

\begin{figure}
	\begin{center}
		\includegraphics[width=\columnwidth]{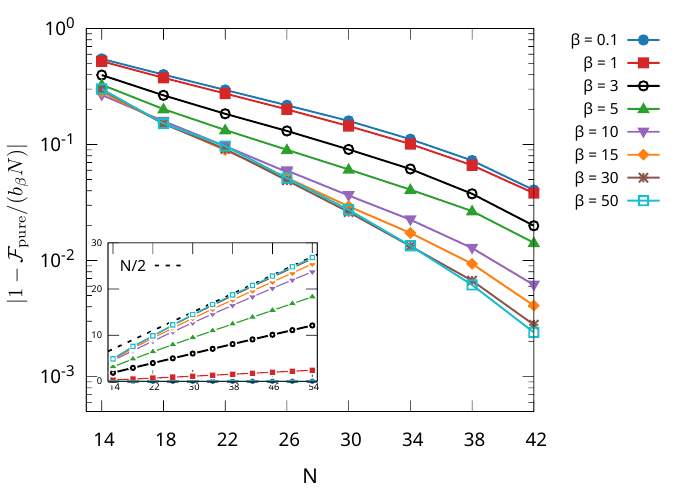}
		\caption{Finite $N$ FAF for the pure KM states for different $\beta$. The leading $N$ correction is $b_\beta N$ with a slope $b_\beta$ that interpolates between $\approx 1/2$ ($\beta \to \infty$) and zero ($\beta \to 0$). The leading correction to the linear growth seems to be exponential but there are deviations in the largest $N$ which prevent us to reach a definite conclusion. Moreover, the correction increases sharply as $\beta$ increases.}
		\label{fig:fafpuredifbeta}
	\end{center}
\end{figure}

We now turn to the finite-$N$ FAF for the pure KM states for different $\beta$. In agreement with Eq.~(\ref{eq:fafbana}), the FAF grows linearly with $N$ in the large $N$ limit, see inset Fig.~\ref{fig:fafpuredifbeta}, with a $\beta$-dependent slope that interpolates between zero (Gaussian states) for $\beta \to 0$, and $N/2$ for $\beta \to \infty$.  This provides a continuous knob to tune quantum magic in a strongly interacting quantum chaotic system. We have also studied the leading finite $N$ correction to this linear scaling. Results depicted in Fig.~\ref{fig:fafpuredifbeta} indicate that the subleading, in $N$, correction to the FAF is orders of magnitude smaller in the $\beta \gg 1$ limit with respect to the small $\beta$ limit. Its precise dependence on $N$ is more difficult to determine though for intermediate values of~$N$ seems exponential with a decay rate that it is sensitive to $\beta$. Deviations for larger $N$ from this exponential behavior may be due to residual error in the linear fitting necessary to extract the leading contribution. 

Finally, we study numerically the finite $N$ time evolution of the FAF. In Fig.~\ref{fig:FAF_time_evolution}~(a), we compare the analytical results of the previous section with finite $N$ numerical calculations. The agreement with the solution of the SD equations is excellent for all times after a simple linear extrapolation ($N \to \infty$ line). The numerical exponential decay rate $\approx 2.11$ exhibited in Fig.~\ref{fig:FAF_time_evolution}~(d) for $N=50$ is close to the analytical one $\sim 2.52$. In order to observe the oscillations, it would be necessary to carry out a finite size scaling analysis on the numerical results and the comparison would be restricted to at most the first oscillations because the small amplitude of the oscillations and numerical uncertainties of the Krylov's method for long times. We note that an important outcome of the numerical and analytical findings is that the typical FAF saturation time is of order $1/J \sim 1$ and therefore independent on $N$. This is in contrast with the time evolution for other quantum many-body systems \cite{sierant2026b} where an $N$ dependence was reported. Further research is required to clarify whether this is a particularity of the SYK model.  We now investigate the FAF for much longer times of the order of the Heisenberg time $\sim e^{N/2}$. At this time scale, the discretization of the spectrum is important so we expect \cite{garcia2026a,cotler2017} correlation functions to develop a ramp-plateau governed by a random matrix like spectral form factor. Since the FAF depends on $G^{\rm{W}}$ Eq.(\ref{eq:faftana}), a similar behavior occurs for the FAF. Numerical results depicted in Fig.~\ref{fig:FAF_time_evolution}~(c) confirm that this is the case.

\emph{Discussion and Conclusions.--} We have employed FAF in order to characterize quantum magic in states constructed from strongly interacting quantum-chaotic Hamiltonians. We have showed analytically that it is possible to change continuously the quantum magic of KM pure states \cite{kour2017} from zero to $N/2$ by tuning $\beta$, the inverse temperature of the blackhole dual. Likewise, the FAF corresponding to the time evolution of a Gaussian state evolved with the SYK Hamiltonian approaches the saturation value $\approx N/2$ exponentially at an $N$-independent time and a rate which is four times the leading Ruelle-Pollicot resonance of the system. 
Numerical results, for~$N\leq 54$, employing GPU acceleration and Krylov-subspace techniques, confirm these results and also enable us to unveil that subleading in $N$ corrections of the FAF for eigenstates of the dense SYK Hamiltonian decay exponentially at a rate that it is sensitive to whether the eigenstates are close to the ground or around the center of the spectrum. The correction of the former is orders of magnitude larger. For a sparse SYK, the main difference is that for states close to the ground this subleading correction is power-law in $N$ instead of exponential. 
Our results are also directly relevant to low dimensional quantum gravity. 
KM pure states in the large $\beta$ limit are dual \cite{kour2017} to a gravity configuration in a near AdS$_2$ background consisting of a black hole with an end-of-the-world particle behind the horizon. SYK eigenstates close to the ground state control Green's functions in the low temperature limit where the dynamics of the SYK model \cite{maldacena2016} and its gravity \cite{maldacena2016b} is governed by the Schwarzian action. 
Therefore, the FAF in these cases is of direct relevance for the characterization of magic in quantum gravity. By contrast, after a time $t \sim 1/J$,  the real time-evolution of Gaussian states results in states with an energy far from the ground state that in principle does not have a gravity dual. This real time-evolution could still have a gravity interpretation if we combine it with Euclidean evolution in the low temperature limit.
Our results provide direct evidence that the FAF of states associated with strongly interacting Hamiltonians can display a rich quantum magic behavior that can differ substantially from that of Haar-random states. 

\acknowledgments
{ 
A. M. G. G thanks Pengfei Zhang for illuminating discussions especially on the relevance of fermionic anti-flatness to characterize magic. We were partially supported by the National Science Foundation of China (NSFC), Individual Grant No.12374138.
}

\bibliography{cooling}

\startendmatter

\section*{End Matter}

\subsection{FAF in the large \texorpdfstring{$q$}{q} limit}
In the main text, we focus on the SYK Hamiltonian with $q = 4$ Eq.~(\ref{eq:sykq4}). Here, we extend the FAF calculation to the SYK model to the $q\gg 1$ limit \cite{maldacena2016} for which analytical results are available while the model is still quantum chaotic. For general $q-$body interactions, the SYK Hamiltonian is given by, 
\begin{equation}
	H_{\rm SYK}^{(q)}=\ii^{q/2}\sum_{1\le i_1<\cdots<i_q\le N} J_{i_1\cdots i_q}\,\chi_{i_1}\cdots \chi_{i_q},
\end{equation}
where the couplings are independent Gaussian random variables with zero mean and variance
\begin{equation}
	\overline{J_{i_1\cdots i_q}^2}=\frac{(q-1)!J^2}{N^{q-1}}.
\end{equation}
Defining $G_{\rm{E}}(\tau) \equiv \sum_{i=1}^{N} \langle \operatorname{T} \chi_i(\tau) \chi_i(0) \rangle_{\beta} / N$ the Euclidean Green's function, by definition Eq.~\eqref{eq:FAF_def}, the $k=1$ FAF of the KM pure state~$\ket{B_{\vec{s}}(\beta)}$ is given by
\begin{align}
	\mathcal{F}(\beta) & = \frac{N}{2} - 4 \sum_{1 \leq m < n \leq N} \bra{B_{\vec{s}}(\beta)} \chi_{m} \chi_{n} \ket{B_{\vec{s}}(\beta)}^2 \\
	& = \frac{N}{2} - 4 \sum_{n = 1}^{N/2} \bra{B_{\vec{s}}(\beta)} \chi_{2n-1} \chi_{2n} \ket{B_{\vec{s}}(\beta)}^2 \\
	& =  N \Biggl[
	\frac{1}{2} - 8 G_{\rm{E}}(\beta/2)^4
	\Biggr] .
\end{align}
In the second step we use the property of the KM pure states such that only the off-diagonal correlators survive~\cite{kour2017} in the large $N$ limit. In the last step we employ the result~\cite{kour2017,Nosaka:2019tcx}
\begin{equation}
\begin{aligned}
	s_{k}\bra{B_{\vec{s}}(\beta)} & \chi_{2k-1}(\tau) \chi_{2k}(\tau') \ket{B_{\vec{s}}(\beta)} \\ 
	& = 2 \ii G_{\rm{E}} \biggl( \tau + \frac{\beta}{2}\biggr) G_{\rm{E}}\biggl(\tau' + \frac{\beta}{2}\biggr) + O(1/N).
\end{aligned}
\end{equation}

\textit{Temperature dependence.} 
In the large $q$ limit, the SYK Green's function in Euclidean time is analytically solvable \cite{maldacena2016} with 
\begin{equation}
	G_{\rm{E}}(\beta/2) = \frac{1}{2} + \frac{1}{q} \log(\cos(\pi v / 2)) + O(1/q) ,
\end{equation}
and the FAF therefore is
\begin{equation}
	\mathcal{F}(\beta) = N \Biggl[
	\frac{1}{2} -  \frac{1}{2} \Biggl(
	1 + \frac{2}{q} \log(\cos(\frac{\pi v}{2}))
	\Biggr)^4 + \dots
	\Biggr] ,
\end{equation}
where $v$ is given by \cite{maldacena2016} $\beta \mathcal{J} \!=\! \pi v / \cos(\pi v / 2)$ with $\mathcal{J} = \sqrt{q} J/2^{\frac{q-1}{2}}$. At low temperature $ \beta \mathcal{J} \gg 1 $, we have the expansion
\begin{equation}
\begin{aligned}
	\frac{1}{N} \mathcal{F}(\beta) = \frac{1}{2} - \frac{1}{2} \Biggl[
			1 + \frac{2}{q}\biggl( &
			\log(\frac{\pi}{\beta \mathcal{J}}) 
			- \frac{2}{\beta \mathcal{J}}
			+ \frac{2}{(\beta \mathcal{J})^2} \\
			& - \frac{8 + \pi^2}{3(\beta \mathcal{J})^3}
			+ O\left((\beta \mathcal{J})^{-4}\right)
			\biggr)
			\Biggr]^4 .
\end{aligned}
\end{equation}
On the other hand, at high temperature we have
\begin{equation}
\begin{aligned}
	\frac{1}{N} \mathcal{F}(\beta) = 
	& \frac{(\beta \mathcal{J})^2}{2 q}
		- \frac{(5q+9)}{48 q^2} (\beta \mathcal{J})^4 \\
	&	+ \frac{\left(94 q^2+225q+90\right)}{2880 q^3} (\beta \mathcal{J})^6 \\
	& + O\left( (\beta \mathcal{J})^8 \right), \quad
	\beta \mathcal{J} \ll 1 . 
\end{aligned}
\end{equation}

\begin{figure}[!t]
  \centering
  \begin{tikzpicture}
    \node[inner sep=0pt, outer sep=0pt] (panelA) {%
      \includegraphics[width=0.9\columnwidth]{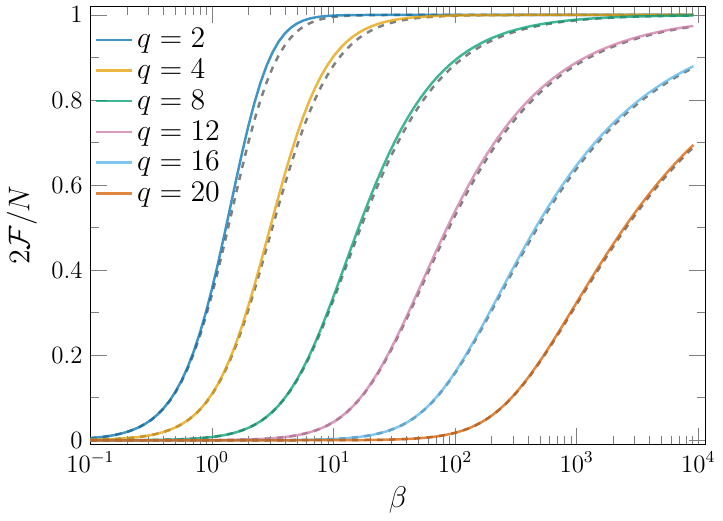}%
    };
    \node[
      anchor=north west,
      inner sep=0pt,
      outer sep=0pt,
      xshift=1pt,
      yshift=-1pt,
      font=\fontsize{9}{10.5}\selectfont\bfseries
    ] at (panelA.north west) {(a)};
  \end{tikzpicture}
  \vspace{-1.5mm}
  \begin{tikzpicture}
    \node[inner sep=0pt, outer sep=0pt] (panelB) {%
     \includegraphics[width=0.9\columnwidth]{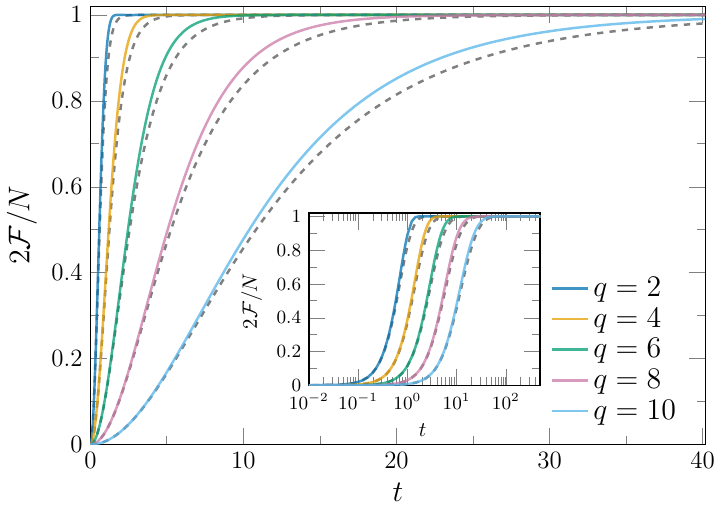}%
    };
    \node[
      anchor=north west,
      inner sep=0pt,
      outer sep=0pt,
      xshift=1pt,
      yshift=-1pt,
      font=\fontsize{9}{10.5}\selectfont\bfseries
    ] at (panelB.north west) {(b)};
  \end{tikzpicture}
  \vspace{-1mm}
  \caption{
	\textbf{(a)} FAF versus $\beta$ for different $q$. \textbf{(b)} Time evolution of the FAF for different $q$ with initial state $\ket{B_{\vec{s}}}$. The gray dashed lines denote the analytic large $q$ solutions and the colored lines denote the solutions of the exact Schwinger-Dyson equations.
  }
\label{fig:FAF_vs_beta_ant_t_qs}
\end{figure}

\textit{Time dependence.}
For the time evolution of FAF with initial states $\ket{B_{\vec{s}}}$, we need to compute the infinite temperature Wightman function defined by $G^>(t) \equiv G^{\rm{W}}(t; \beta = 0) = (N Z_{\beta=0})^{-1} \sum_{j=1}^{N} \tr(\chi_j(t) \chi_j(0))$, which in large $q$ case is
\begin{equation}
\begin{aligned}
	G^>(t) & = \frac{1}{2} \!\sgn(t)\! \Biggl[
	1 - \frac{2}{q} \log(\cosh(\mathcal{J} t)) + \dots
	\Biggr] \\
	& \approx \frac{\sgn(t)}{2 \cosh(\mathcal{J} t)^{\frac{2}{q}}} .
\end{aligned}
\end{equation}
The FAF then is 
\begin{equation}
	\begin{aligned}	
		\frac{1}{N} \mathcal{F}(t) 
		& = \frac{1}{2} - \frac{1}{2} \Biggl[
		1 - \frac{2}{q} \log(\cosh(\mathcal{J} t)) + \dots
		\Biggr]^{4} 
		\\
		& \approx \frac{1}{2} - \frac{1}{2 \cosh(\mathcal{J} t)^{8/q}} .
	\end{aligned}	
\end{equation}
The short time expansion is
\begin{equation}
	\begin{aligned}
		\frac{1}{N} \mathcal{F}(t) = &
		2^{2-q} (J t)^2 
		- \frac{1}{3} 4^{1-q} (q+9) (Jt)^4 \\
		&+\frac{1}{45} 8^{1-q} \left(4 q^2+45 q+90\right) (Jt)^6
		+O\left((Jt)^8\right) .
	\end{aligned}
\end{equation}
Moreover, from the exponentiated form of the Green's function, the quasi-normal modes of FAF are 
\begin{equation}
	\omega_n = - \ii \mathcal{J} \biggl(
	2 n + \frac{8}{q}
	\biggr) ,
	\quad
	n = 0, 1, 2, \dots
\end{equation}
This gives the long time decay rate $8 \mathcal{J}/q$ of FAF in the large $q$ limit. The plot of $\mathcal{F}(t)$ for different $q$ is presented in Fig.~\ref{fig:FAF_vs_beta_ant_t_qs}~(b).

\end{document}